# Strange and nonstrange sea quark-gluon effects in nucleons

M. Batra[a], A. Upadhyay[a]

[a]School of Physics and Material Science,
Thapar University, Patiala, Punjab -147004

Probabilities of various Fock states with strange and nonstrange quark-gluon sea contents are calculated to probe the hadronic structure. Particularly for nucleon, we find various contributions to the low energy properties from scalar, vector and tensor sea in addition to three valence quarks. We focus on the importance of individual sea contributions to the low energy parameters of nucleon by taking the strange and non strange quark-gluon content to the hadron sea. We confirm that the extended Fock space wavefunction is capable of explaining the experimental results where vector sea plays a crucial role in studying hadronic structure while scalar and tensor sea appears to be less dominating due to quark-spin flip process but cannot be neglected. Some of the properties like spin distribution and $g_A/g_V$ ratio seem to be the most affected by the change in the statistically determined coefficients. Detailed analysis includes different approximations within the statistical approach to test the validity of the model chosen. Phenomenological implication of such sea affecting these properties is also discussed and the results are compared with the experiments.

# I. Introduction

QCD is a theory that successfully describes hadronic phenomenon at very short distance scales by describing the interaction between quarks through exchange of gluons. But at long distance scales, due to non-perturbative nature of interaction of quarks and gluons, many of fundamental particles and their internal structure is still a mystery. In particular, to bridge the gap between QCD and current description of hadron structure, discrete models are suggested in literature. These models describe baryon to be composed of three valence quarks and sea filled with gluons and quark-antiquark pair as condensates. The effect of sea contributions to nucleonic properties like weak decay coupling ratios, magnetic moment, masses can be studied via these models. The spin structure of proton is one of the key examples confirming the presence of gluons by providing a non negligible contribution to spin of proton. The latest studies show that only 30% of the total spin of nucleon is carried by quarks as compared to the rest part including gluons. Many phenomenological models have contributed their studies to solve these kinds of puzzles [1-3, 17].

Static properties of the hadron is studied through variety of experiments [4-6]. The well known experiments are EMC (Electron-Muon Collaboration) and SMC (Spin-Muon Collaboration). Wide range of data is also available from experiments at SLAC [7-9]. However, recent experiments show that strange quark also makes a significant contribution to nucleonic spin that can be studied using polarized deep-inelastic scattering experiments of electrons or muons from nucleons. NuTeV Collaboration at FermiLab predicted the non-zero value of their contribution to spin of nucleons [10-11] via strange quark content ratio which is the fraction of nucleon momentum carried by strange quark to non-strange quark as $\frac{2(s+\bar{s})}{(u+\bar{u}+d+\bar{d})} = 0.477 \pm 0.063 \pm 0.053$ [12]. Value of this ratio points to the existence of strange quark in the sea. The complete information about strange quark effects in the hadronic sea is still being explored and hence we must check the authenticity of the models in retrieving the experimental observations regarding the hadron structure. Despite many experiments, the spin content of nucleon and deuteron is not understood well. At Present we have several phenomenological models in hand to justify the experimental data for looking into the spin distribution, magnetic moment and axial coupling constants.

In this paper we study the statistical model and simple quark model to look for the low energy parameters. Models discussed here assume the proton structure to be made up of two parts one is valence part and other is virtual sea which has the quark-antiquark pair multi-connected through gluons. A suitable wave-function encloses spin, flavor and color for valence and sea-part to produce quantum number of baryon being framed and brief formalism is described in section II. Section III discusses the fundamental properties at low energies and their dependence on various coefficients mentioned in the formalism. Discussion and results will be followed with tabulated results in section IV. Concluding remarks is given in section V.

## II. Wave-function for Nucleons

A spin up nucleon state is expanded in terms of quark-gluon Fock states consisting of three valence quarks in terms of wavefunction $\Phi$ and a sea consisting of quarks, antiquarks and gluons, containing maximum five constituents having definite spin as H and color as G. The expansion of Fock states in spin and color space is done using the assumption of equal probability for each sub state. We also use the approximation wherein a quark in the core is not antisymmetrized with its identical flavor in the sea, and have treated quarks and gluons as non-relativistic particles moving in S-wave motion. Sea with spin 0,1,2 as $H_{0,1,2}$ and color as $G_{1,8,\overline{10}}$ wave function satisfy following relations:

$\langle H_i | H_j \rangle = \delta_{ij}$, $\langle G_k | G_l \rangle = \delta_{kl}$

Two gluons, each with spin 1 combine to give spin as 0, 1 or 2 as below:

Spin: gg: $1 \otimes 1 = 0s \oplus 1a \oplus 2s$

Similarly for color space we have

Color: gg: $8 \otimes 8 = 1s \oplus 8s \oplus 8a \oplus \overline{10}a \oplus \overline{10}a \oplus 27s$.

Similar treatment is extended for the multiple gluon cases upto three. The maximum constituent including quark and gluon will be five. Nucleon wave function can be written with all the possible combinations that can give the proton a spin ½, flavor octet and color singlet as:

$\Phi_1^{\frac{1}{2}}H_0G_1$, $\Phi_8^{\frac{1}{2}}H_0G_8$, $\Phi_{10}^{\frac{1}{2}}H_0G_{\overline{10}}$, $\Phi_1^{\frac{1}{2}}H_1G_1$, $\Phi_8^{\frac{1}{2}}H_1G_8$, $\Phi_{10}^{\frac{1}{2}}H_1G_{\overline{10}}$ and $\Phi_8^{\frac{3}{2}}H_1G_8$, $\Phi_8^{\frac{3}{2}}H_2G_8$

We assume proton to have spin 1/2, color singlet and flavor octet. Here the valence quarks with spin 3/2 and color octet $\Phi_8^{\frac{3}{2}}$ wave-function can give spin ½ to the proton, if the sea quark-gluon is having spin either 1or 2. All other possibilities like $H_2G_1, H_2G_{\overline{10}}$ are excluded as they will not give rise to color singlet state. Contributions from states like $H_0G_{27}$, $H_2G_{27}$ are ignored due to the suppressed higher multiplicities.

Here a single gluon in the sea contributes to the sea with spin 1 and color octet, i.e $H_1G_8$ only. The contributions from $H_0G_{\overline{10}}$ and $H_1G_1$ sea for two gluon is excluded, since $H_0$ and $G_1$ are symmetric whereas $H_1$ and $G_{\overline{10}}$ are anti-symmetric under the exchange of two gluons making the wave-function anti-symmetric which is unacceptable for the bosonic system.

The total flavor-spin-color wave function of a spin up baryon:

$$|\Phi_{\frac{1}{2}}^{\uparrow}\rangle = \frac{1}{N}[\Phi_1^{(\frac{1}{2})^{\uparrow}}H_0G_1 + a_8\Phi_8^{(\frac{1}{2})^{\uparrow}}H_0G_8 + a_{10}\Phi_{10}^{(\frac{1}{2})^{\uparrow}}H_0G_{\overline{10}} + b_1[\Phi_1^{\frac{1}{2}}\otimes H_1]^{\uparrow}G_1 + b_8(\Phi_8^{\frac{1}{2}}\otimes H_1)^{\uparrow}G_8 + b_{10}(\Phi_{10}^{\frac{1}{2}}\otimes H_1)^{\uparrow}G_{\overline{10}} + c_8(\Phi_8^{\frac{3}{2}}\otimes H_1)^{\uparrow}G_8 + d_8(\Phi_8^{\frac{3}{2}}\otimes H_2)^{\uparrow}G_8]$$

Where $N^2 = 1 + a_8^2 + a_{10}^2 + b_1^2 + b_8^2 + b_{10}^2 + c_8^2 + d_8^2$

Here N is the normalization constant and scalar, vector and tensor sea contribution of the nucleon is coming in terms of $a_0, a_8, a_{10}, b_1, b_8, b_{10}, c_8, d_8$ coefficients that can give the properties in terms of alpha and beta as given in Table 2. An operator formalism can also be used to retrieve information about the hadrons.

In this formalism, spin and flavor operator are suitably switched between two eigenstates in the form $\hat{O} = \sum_i \hat{O}_f^i \hat{\sigma}_z^i$ where $\hat{O}_f^i$ depends upon on the flavor of ith quark and $\hat{\sigma}_z^i$ is the spin projection operator of $i^{th}$ quark and $\langle\phi^\rho|O_f^i|\phi^\rho\rangle$ and $<\hat{\sigma}_z^i>^{\rho\uparrow\rho\uparrow} = \langle\chi^{\rho\uparrow}|\sigma_z^i|\chi^{\rho\uparrow}\rangle$, $<\hat{O}_f^i>^{\lambda\rho} = \langle\phi^\lambda|O_f^i|\phi^\rho\rangle$, $<\hat{\sigma}_z^i>^{\lambda\uparrow\rho\uparrow} = \langle\chi^{\lambda\uparrow}|\sigma_z^i|\chi^{\rho\uparrow}\rangle$ etc. where $\lambda$ denotes the symmetric wave-function and $\rho$ denotes the antisymmetric wave-function. Thus the modified wave function is then expressed in terms of parameters defined below:

$a=\frac{1}{2}(1-\frac{b_1^2}{3})$, $b=\frac{1}{4}(a_8^2-\frac{b_8^2}{3})$, $c=\frac{1}{2}(a_{10}^2-\frac{b_{10}^2}{3})$, $d=\frac{1}{18}(5c_8^2-3d_8^2), e=\frac{\sqrt{2}}{3}b_8c_8$.

The above defined parameters can be further be related to the coefficients α and β as:
$\alpha = \frac{1}{N^2}(\frac{4}{9})(2a+2b+3d+\sqrt{2}e)$
$\beta = \frac{1}{N^2}(\frac{1}{9})(2a-4b-6c-6d+4\sqrt{2}e)$

The properties like magnetic moment, spin distribution, axial coupling constant ratios are calculated using nucleon wave function at low energies typically of 1 GeV scale with the help of the coefficients defined above.

Here our aim is to study the proton consisting of three quarks in core and quark-antiquark pairs and gluons in the sea to form overall antisymmetrization for the proton and represented by wave function
$|p\rangle = \sum_{i,j,k}C_{ijk}|uud,i,j,k\rangle$
Where i is the number of $u\bar{u}$ pair, j is the number of $d\bar{d}$ pair and k is the number of gluons. The probability of finding the proton in Fock state $|uud,i,j,k\rangle$ is: $\rho_{i,j,k} = |C_{i,j,k}|^2$ and satisfy the condition of normalization as:

$$\sum_{i,j,k}\rho_{i,j,k} = 1.$$

All the probability ratios are determined by using principle of detailed balance without any parameter. The principle here assumes the expansion of hadrons in terms of quark-gluon Fock states where Fock states assume the presence of quark-antiquark pairs multi-connected to gluons and exchange between any two Fock states balance each other. This principle proposed by Zhang et al. [13-15] has been successful in explaining the flavor asymmetry of hadrons and Parton distribution function of proton.[16] This principle calculates $\bar{d}-\bar{u} = 0.124$ in close agreement with the experimental data 0.118±0.01 [23]. Recent applications of this principle include the computation of parton distribution function [24]. Thus, it is clear that such type of Fock states having definite spin and quantum numbers and hence symmetry properties can be used to find quark distribution of spin among nucleons. We calculated the probability of such Fock state including strange quark in the sea, extending the work given by Zhang et al. [13] and utilize it to find probability ratios in spin and color spaces. Based on the probabilities given in Ref.[17] we find the additional multiplicities from Fock states like $\overline{uuuuuu}, \overline{dddddd}, \overline{uudd}\overline{ss}$ assuming strange quark to be an essential component of sea-part. The relative probability in spin and color space is taken where the core part should have an angular momentum $j_1$ and sea have $j_2$ and the total angular momentum should come out as ½ (may be $j_{1+}$ $j_{2=}$½ ) with a color singlet state. Relative probabilities thus expressed in the form of a common multiplier "c" and computed in the form of "nc" , where n represents the suitable multiplicity for that Fock state. This will contains the information about the Fock states present in the sea, with their relative probabilities in color and spin space. The sum of the total probabilities will give the coefficients $a_0$, $a_8$, $a_{10}$, $b_1$, $b_8$, $b_{10}$, $c_8$, $d_8$ to the total wavefunction. Here scalar, vector and tensor sea contribution of the nucleon is coming in terms of these coefficients that can give the properties in terms of alpha and beta. We here apply few modifications to our calculations by assuming the contributions from three gluon states to be zero for $H_0 G_{\overline{10}}$ and $H_1 G_{\overline{10}}$ and for the states which consider the spin of valence part to be 3/2 forming a color singlet state only.

### Strange Quark Content in Nucleon

Recent experimental and theoretical studies [12,18] show that the strange quark is one of the essential component of intrinsic sea that contributes to spin distribution among quarks and gluons within nuclei. Thus we extend our original work by including Fock states with $\overline{ss}$. The Fock states without $\overline{ss}$ covers 86% of the total Fock states while inclusion of strange quark content reduces it to 80%. Moreover, if we assume sea to have strange quark then its mass should be taken into consideration through the proper constrain.

To accommodate the strange quark, and to have processes like $g \Leftrightarrow \overline{ss}$ we should have the system with energy that is larger than two times that of the proton mass. For this, we first extend principle of detailed balance to get in hand the probability of each Fock state which include $\overline{ss}$ content in Fock states with due consideration of mass of strange quark. The generation of $\overline{ss}$ pair from gluons is restricted by applying a constraint in the form of $k(1-C_l)^{n-1}$ [16] where $n$ is the total number of partons present in the Fock state. The factor comes as a result of gluon free energy distribution and constraints to the momenta and total energy of partons present in the proton. In all cases $C_{l-1} = \frac{2M_s}{M_p - 2(l-1)M_s}$, $M_s$ is the mass of s-quark and $M_p$ is the mass of proton. The strange quark-antiquark distributions in the nucleon limits the number of strange quark-antiquark pair. According to the distribution, only one $\overline{ss}$ pair can be accommodated at time. To check the validity of our assumptions, flavor asymmetry and strange quark content ratio are also estimated and matched with results available.

### III. Nucleonic Parameters

Low energy properties of nucleonic system can be estimated in various models, particularly in SU(6) based Naïve Quark Model. It explains magnetic moment of all baryons to a better extent rather than spin distribution among baryons. Although experiments and phenomenologist try to solve the puzzles regarding spin distribution but the complete picture is not clear.

Most of the properties can be calculated by integration of g1 over x and called first moment: $\Gamma_1 = \int_0^1 g1(x)dx = \frac{1}{2}\int_0^1 \sum_f e_f^2 \Delta q_f(x)dx$

The modified axial representation is:
$$\Gamma^1(p,n) = \pm\frac{1}{12}(\Delta u - \Delta d) + \frac{1}{36}(\Delta u + \Delta d - 2\Delta s) + \frac{1}{9}(\Delta u + \Delta d + \Delta s)$$
$$= \frac{1}{12}(\pm a_3 + \frac{1}{\sqrt{3}}a_8) + \frac{1}{9}a_0$$

Here $\Delta u, \Delta d, \Delta s$ is to be calculated to get nucleonic properties.

## A. Spin Distribution for partons:

Nucleon spin is said to be distributed among valence and sea quarks and gluons and some of the part is carried by their orbital angular momenta. Polarized Deep Inelastic scattering experiments is the key tool for probing the internal structure of nucleon. The measurement of spin distribution function was given by EMC experiment in 1988 and measured $\Gamma_p^1$ as 0.126±0.018 [18]. Experimentally, SMC (Spin Muon Collaboration) [19] predicted very less contribution from intrinsic spin of quarks to proton spin. COMPASS and HERMES collaborations focused on the measurement of $\Delta\Sigma$ [20] [21] at $Q^2$ =3 GeV$^2$ and 5 GeV$^2$ respectively. The total contribution of spin among quarks inside the proton can be measured through first moment of structure function $g_1$ which can further be expressed in terms of polarized quark and anti-quark distribution. QCD corrections involve the $Q^2$ dependence that replaces $g_1$ by $g_1(x, Q^2)$. A next to leading order corrections to $g_1$ upto third order is given by J. Ellis [22]. The spin structure function $\Gamma_p^1$ is also important, as this could further be utilized to obtain the three matrix elements $a_0$, $a_3$ and $a_8$ under SU(3) flavor symmetry. It is also written in terms of polarized quark and anti-quark densities. The matrix element $a_0$ gives the contribution $\Delta\Sigma = (\Delta u + \Delta d + \Delta s)$. Spin distribution also includes gluonic spins and momentum of quarks and gluons as per helicity sum rule [19]. In our models, charged squared spin projection operator $I_1^p = \frac{1}{2} < \sum_i e_i^2 \sigma_z^i >_p$ and $I_1^n = \frac{1}{2} < \sum_i e_i^2 \sigma_z^i >_n$ produce $I_1^p = \frac{1}{6}(4\alpha - \beta)$ and $I_1^n = \frac{1}{6}(\alpha - 4\beta)$. The operator here provides contribution in terms of quark and gluon Fock states. Furthermore, the orbital angular momentum share is not considered here due to very less overlap regions between momentum of valence and sea [18].

## B. Axial Coupling Constant Ratio and Matrix Elements:

The matrix element of quark currents between proton and neutron states in beta decay may be calculated using isospin symmetry. From the semileptonic decay B→B`e $\nu$, vector and axial coupling constants $g_v$ and $g_A$ can be determined. The matrix element $a_3$ represents weak decay coupling constant ratio for proton and neutron. Operator for this ratio can be taken as: $\widehat{O}_f^i = 2I_3^i$ and this gives $g_A/g_V = 3(\alpha + \beta)$ for neutron decaying into proton. Moreover, QCD Corrections gives

$$\Gamma_1(Q^2) = \int_0^1 g1(x,Q^2)dx = \frac{1}{18}(4\Delta u + \Delta d + \Delta s)(1-\alpha_s(\pi))$$

where $(1-\alpha_s\pi)$ is the first order corrections derived from Bjorken sum rule [22]. On the basis of SU(3) symmetry, all baryon decay rates depend upon two universal parameters F and D. The ratio of F and D is interpreted as $\frac{F}{D} = \frac{\alpha}{\alpha+2\beta}$. Experimentally, F and D have been determined from hyperon $\beta$-decays as F=0.463±0.08 and D=0.804±0.08 [23]. In general, F and D can be related to SU(3) group structure constants.

## C. Magnetic Moment Ratio

Magnetic moment of all baryons can be related to spin distribution of quarks ($\Delta u, \Delta d, \Delta s$) without giving explicit wavefunctions. It can be written as:
$\mu_B = \sum_q (\Delta q)^B \mu_q$ (q=u,d,s). Gupta et al. [24] determined the magnetic moments of all baryons using these expression as $\mu_p = 2.792, \mu_n = -1.913$ in their three and four parameter fits.

In our model, magnetic moment ratio of proton and neutron is described either in terms of quark magnetic moment operator or in terms of two parameters $\alpha$ and $\beta$ as $\frac{\mu_P}{\mu_N} = -\frac{2\alpha+\beta}{\alpha+2\beta}$ calculated from probabilities. The operator for magnetic moment ratio of proton and neutron is given as $\hat{\mu} = \sum_q \frac{e_q^i}{2m} \sigma_z^q, q = (u,d,s)$ where $\mu_p = 3(\mu_u \alpha - \mu_d \beta)$ and $\mu_n = 3(\mu_d \alpha - \mu_u \beta)$.

## IV. Discussion and Results

We study two models assuming sea containing admixture of gluons and quark-antiquark pair in addition to the three valence quarks. The comprehensive analysis is based on the simple quark model and different approaches within the statistical models named C, P and D as discussed below.

Model C, P and D within statistical model have been studied to develop a better understanding in distinguishing these approach based on experimental data. All the low energy properties mentioned above are calculated in model C, P and D with two different sets of Fock states, one is with $s\bar{s}$ and other is without $s\bar{s}$. We also elaborate our discussion by comparing the result coming from distinct scalar, vector and tensor sea.

Model C is the basic model having various quark-gluon Fock states in which the three quark core and the rest of the sea have definite spin and color quantum numbers, using the assumption of equal probability for each sub state of such a state of the nucleon[3].

Model P is based on the constrain where quark-antiquark pair can reside in the form of colorless pseudo-scalar Goldstone bosons which may appear due to gluonic exchange interactions and spin-flip process. These internal pseudoscalar bosons lead to some more symmetry in quark-gluon Fock states and sea is further no more assumed as an active participant in the nucleons. In case of $|gg\ q\bar{q}\rangle$ state, in order to compensate the odd parity of the qq pair, one of the gluons will be assumed to be in TE mode while the other in TM mode

Model D here comes with a modification of suppression of Fock states with higher multiplicities. Sea with larger color multiplicity has less probability of survival due to larger possibility of interaction. Suppression is here done by dividing the probability factors involved in model 'C' by multiplicities in spin and color space. The relative probabilities and the respective numbers from these approaches is shown in table 1.

The statistical approach is based on the principle of balance and detailed balance [14]. Model assumes proton to be made up of different Fock states and the probability associated with each Fock states is further utilized to compute low energy properties in terms of the coefficients $\alpha$ and $\beta$. From α and β, all the low energy parameters are calculated which can finally be written in terms $a_0, a_8, a_{10}, b_1, b_8, b_{10}, c_8, d_8$ coefficients as given in Table 2.The statistical model and its approaches are used to find the separate contributions from scalar, vector and tensor sea. The quark model[3] result is also modified with the inclusion of the so called negligible contributions where the non zero contributions from scalar and tensor sea are taken from the statistical approach. The new coefficients obtained without $s\bar{s}$ and with the strange quark condensates within these approach are shown in table 3.

On the other hand, a simple quark model [3] computes $\alpha$ and $\beta$ analytically by considering vector sea to find a parameter set using the data on magnetic moment and weak decay coupling constant and compute the coefficients by simple fitting of these parameter. The magnetic moments of all the baryons can be expressed in terms of quark magnetic moment, $\alpha$, and $\beta$ as defined in section III. Using these α and β, an appropriate parameter set for the coefficients contributed by vector sea is reproduced. Here the vector sea is considered as only active contributor from the total sea. The vector sea dominance comes in the form of just four fitted parameters as $b_1=0.0642$, $b_8=0.47$, $c_8=0.16$, $a_{10}=0.3$. Motivation to neglect tensor sea emerges from the fact that tensor sea contribution comes from spin 3/2 valence part and it becomes less probable for core part to have spin 3/2. Here most of the scalar sea contribution is suppressed to the fact that they can only come from two gluon sea and this approximation led result close to experimental values.

The sea contributions in different models and their approximated version is presented in table to check the validity of these models, and matched with experimental results and thereby showing the importance of vector, tensor and scalar sea in finding these properties. Each term has non negligible dependence on almost all the nucleonic parameters. Here to check the contribution from the scalar sea, we suppress the vector and tensor sea contributions and similar approach to find the individual contribution from vector and tensor sea. As the sea part is dominated by emission of virtual gluons so we can expect $b_8$ and $c_8$ to be more dominant. If only vector sea is assumed to be contributing, then nucleonic properties like coupling constant and F/D ratio are mainly affected with parameters $b_8$ and $c_8$.

It can be seen from Table 3 that for simple quark model, if we take non zero scalar and tensor sea contribution from the available statistical data, then the percentage error increase upto 7-8% except spin distribution to the nucleons. Here the deviation goes down from ∼58 % to 6-7 % approximately. Tensor sea appears to be less dominating due to quark-spin flip process but cannot be neglected in all cases. Vector sea plays an important role in determining these values close to experiments. Some of the properties like spin distribution and $g_A/g_V$ ratio seem to be the most affected by the change in the values of these coefficients. We tabulated the result showing extent to which the sea contribution is affecting the nucleon properties in table 3.

For instance, neglecting the scalar and tensor sea in statistical model (C, P, D) and simple quark model, the magnetic moment ratio deviates by 6-10% as compared to experimental data[27] shown in table 3. Weak decay matrix element ratio deviates by 30-40% from original value when the sea contribution is included in statistical model as well as in simple quark mode. But when sea is excluded here the ratio is pretty close to the experiments where deviation is 0-17% which is much smaller than the earlier case. F/D ratio is more close to experimental data when sea is excluded in all the above approach but when the sea is taken into account then statistical approach give much better results especially for C model than other and SQM. As we go from simple quark model to statistical, the sea become the dominant contributor for the spin distribution and therefore the results are better matched with experiments in later case. Similar observation holds for the above conclusion when we go from exclusion of sea to the inclusion of sea in C model.

Similarly, inclusion of strange quark condensates in sea produce results more close to experimental predictions. The extension of principle of detailed balance is checked against $\bar{u}/\bar{d}$ and $\bar{u}-\bar{d}$ asymmetry with a value 0.71 for $\bar{u}/\bar{d}$ and 0.124 for $\bar{u}-\bar{d}$ asymmetry [25] which is found to be matching well with the experiments and other theoretical models. The principle here finds the values of strange quark content ratio to be

$$\frac{2\bar{s}}{(\bar{u}+\bar{d})} = 0.37 \; ; \; \frac{2\bar{s}}{(u+d)} = 0.03$$

Matching with results from NuTeV Collaboration[12].

## V. Conclusion

In our work nucleon is assumed to have a virtual sea with strange and nonstrange quark condensates in addition to gluons. Probabilities for the nucleon to have sea consisting of limited number of quark condensates and gluon are calculated taking the fact that gluons if present are capable of annihilating to these condensates. The two parameters α and β directly relates these probabilities to sea contributions in terms of five coefficients representing scalar, vector and tensor sea. The physical significance of two parameters also lies in their relation with number of spin up and spin down quarks in the spin up baryon that is $\Delta q = n(q \uparrow) - n(q \downarrow) + n(\bar{q} \uparrow) - n(\bar{q} \downarrow)$, $q = u, d, s$. $\Delta q$ is used to find the spin structure of proton, weak decay coupling constant ratios and the magnetic moment of baryons. Our main attention is to find the impact of strange and non strange partons on the spin distribution and magnetic moment of nucleon. Our calculations hold good for scale of the order of 1GeV$^2$ whereas experimental results are applied at typical energy scale $Q^2 \sim 10$ GeV$^2$. In a closure view, on neglecting the scalar and tensor sea completely, all the properties are not retrieved simultaneously, but their inclusion gives better results for magnetic moment and spin distribution for proton and neutron. We conclude that inclusion of scalar and tensor sea is important so as to produce more accurate results close to experiments.

Table 1: Various Fock states and contribution to their relative probabilities in model C and D

| States | Without $s\bar{s}$ | | | With $s\bar{s}$ | | |
|---|---|---|---|---|---|---|
| | Relative probability | Value of c | Value of d | Relative probability | Value of c | Value of d |
| $|gg\rangle$ | 0.081887 | 0.005118 | 0.03887 | 0.070302 | 0.00439 | 0.03374 |
| $|\bar{u}ug\rangle$ | 0.054978 | 0.001718 | 0.01900 | 0.009118 | 0.0028 | 0.0031516 |
| $|\bar{d}dg\rangle$ | 0.08270 | 0.002585 | 0.028586 | 0.013678 | 0.0004274 | 0.004727 |
| $|\bar{s}sg\rangle$ | ------ | ------ | ------- | 0.09582 | 0.00994 | 0.004727 |
| $|\bar{u}u\bar{d}d\rangle$ | 0.02930 | 0.000916 | 0.010129 | 0.01597 | 0.000499 | 0.005519 |
| $|\bar{d}d\bar{d}d\rangle$ | 0.01457 | 0.000910 | 0.006916 | 0.03194 | 0.01368 | 0.003790 |
| $|ggg\rangle$ | 0.03205 | 0.004579 | 0.02358 | 0.044768 | 0.00639 | 0.0037903 |
| $|\bar{u}u\bar{u}u\bar{u}u\rangle$ | ------ | ------ | ------- | 0.000266 | 0.000038 | 0.0001958 |
| $|\bar{d}d\bar{d}d\bar{d}d\rangle$ | ------ | ------ | -------- | 0.000665 | 0.000049 | 0.000494 |
| $|\bar{u}u\bar{d}dg\rangle$ | 0.03057 | 0.000477 | 0.012858 | 0.002761 | 0.000995 | 0.001161 |
| $|\bar{u}u\bar{s}sg\rangle$ | ----- | ----- | ------- | 0.03194 | 0.000499 | 0.020151 |
| $|\bar{d}d\bar{s}sg\rangle$ | ------- | ------ | ------- | 0.04791 | 0.000749 | 0.001580 |
| $|\bar{d}d\bar{d}dg\rangle$ | 0.01524 | 0.000127 | 0.003224 | 0.01380 | 0.000115 | 0.006752 |
| $|\bar{u}ugg\rangle$ | 0.03042 | 0.000253 | 0.01900 | 0.014193 | 0.000118 | 0.030007 |
| $|\bar{d}dgg\rangle$ | 0.04544 | 0.000379 | 0.009607 | 0.02129 | 0.0001774 | 0.045010 |
| $|\bar{s}sgg\rangle$ | ------ | ----- | ------ | 0.052265 | 0.0004356 | 0.011049 |
| $|\bar{u}u\bar{u}u\rangle$ | 0.00723 | 0.0004519 | 0.000343 | 0.003992 | 0.009118 | 0.001895 |
| $|\bar{u}u\bar{u}ug\rangle$ | 0.00764 | 0.0000368 | 0.001616 | 0.000691 | 0.0000057 | 0.00014519 |
| $|\bar{u}u\bar{d}d\bar{s}s\rangle$ | ------ | ------ | ----- | 0.015197 | 0.001478 | 0.054700 |

Table 2: Computed values of coefficients in three models and in two forms (with $s\bar{s}$ & without $s\bar{s}$) are:

| Sr. No. | Coefficients | Statistical Model without $s\bar{s}$ | | | Statistical Model with $s\bar{s}$ | |
|---|---|---|---|---|---|---|
| | | C | P* | D | C | D |
| 1. | $a_0$ | 1 | 1 | 1 | 1 | 1 |
| 2. | $a_8$ | 0.517 | 0.470 | 0.2017 | 0.972 | 0.23 |
| 3. | $a_{10}$ | 0.0825 | 0.147 | 0.0759 | 0.402 | 0.0912 |
| 4. | $b_1$ | 0.1201 | 0.206 | 0.467 | 0.494 | 0.607 |
| 5. | $b_8$ | 1.760 | 1.119 | 0.050 | 1.579 | 0.0588 |
| 6. | $b_{10}$ | 0.1984 | 0.261 | 0.0641 | 0.589 | 0.073 |
| 7. | $d_8$ | 0.8503 | 0.792 | 0.055 | 0.679 | 0.074 |
| 8. | $c_8$ | 0.2439 | 0.299 | 0.3494 | 1.092 | 0.406 |

*Model 'P' assumes $q\bar{q}$ as inactive contributor thus inclusion of strange quark does not modify the results in this case of strange quarks.

Table 3- Comparison of the calculated magnetic moment ratio, spin distribution and weak decay coupling constant for nucleons in different approaches(C,P,D) of the statistical model, with the parameters obtain in the simple quark model.

| Parameter | Statistical model with all sea | | | Statistical model without scalar and tensor | | | Quark model with all sea* | | | Quark Model without scalar and tensor sea | Statistical Model with strange sea(at most one $s\bar{s}$) | | Experimental Results |
|---|---|---|---|---|---|---|---|---|---|---|---|---|---|
| | Models with modifications | | | Model with modifications | | | Model with modifications | | | | | | |
| | C | P | D | C | P | D | C | P | D | | C | D | |
| $\alpha$ | 0.216 | 0.297 | 0.299 | 0.255 | 0.2853 | 0.3778 | 0.2141 | 0.29912 | 0.3284 | 0.3415 | 0.19 | 0.259 | |
| $\beta$ | 0.0715 | 0.0808 | 0.0818 | 0.0917 | 0.1003 | 0.1075 | 0.03955 | 0.05876 | 0.0728 | 0.07749 | 0.0607 | 0.0732 | |
| $\mu_p/\mu_n$ | -1.402 | -1.47 | -1.46 | -1.37 | -1.38 | -1.46 | -1.59 | -1.57 | -1.55 | -1.53 | -1.41 | -1.46 | -1.46 [28] |
| $g_A/g_V$ | 0.863 | 1.133 | 1.143 | 1.04 | 1.16 | 1.45 | 0.7609 | 1.074 | 1.20 | 1.26 | 0.744 | 1.02 | 1.257±0.03 [30][26] |
| F/D | 0.611 | 0.647 | 0.646 | 0.581 | 0.587 | 0.637 | 0.730 | 0.718 | 0.693 | 0.688 | 0.607 | 0.639 | 0.575±0.016 [30][26] |
| $I_1^p$ | 0.132 | 0.184 | 0.185 | 0.155 | 0.173 | 0.233 | 0.136 | 0.189 | 0.207 | 0.215 | 0.115 | 0.160 | 0.127±0.004 [29] |
| $I_1^n$ | -0.011 | -0.043 | -0.047 | -0.018 | -0.019 | -0.008 | 0.009 | -0.010 | 0.006 | 0.0052 | -0.010 | -0.0064 | -0.030 [29] |